\documentclass[aps,prl,reprint,superscriptaddress,citeautoscript,longbibliography,floatfix]{revtex4-2}

\usepackage{amssymb}
\usepackage{amsmath}
\usepackage{amsfonts}
\usepackage{bm}

\usepackage{graphicx}
\usepackage{tikz}
\usepackage{tikz-3dplot}
\usetikzlibrary{math}
\usetikzlibrary{arrows.meta}

\usepackage{algorithm}
\usepackage{algpseudocode}

\usepackage{physics2}
\usephysicsmodule{ab}
\usepackage{derivative}
\newcommand{\ve}[1]{\bm{#1}} % 2 vector symbols: \vec and \ve

\newcommand{\Rset}{\mathbb{R}}

\usepackage{cprotect}
\usepackage{comment}
\usepackage[normalem]{ulem}
\usepackage{xcolor}

\usepackage{xr-hyper}
\usepackage{hyperref}
\usepackage[capitalise]{cleveref}
\usepackage[color=yellow]{pdfcomment}

\newcommand{\pseudoparagraph}[1]{\relax}
\newcommand{\pseudosubparagraph}[1]{\relax}
\renewcommand{\pseudoparagraph}[1]{\textit{#1}}
\renewcommand{\pseudosubparagraph}[1]{\textit{#1}}

\begin{document}

\title{Symmetry-Informed Term Filtering for Continuum Equation Discovery}

\author{Junya Yokokura}
\email{junya-yk@g.ecc.u-tokyo.ac.jp}
\affiliation{Department of Physics,\! The University of Tokyo,\! 7-3-1 Hongo,\! Bunkyo-ku,\! Tokyo 113-0033,\! Japan}%

\author{Kazumasa A. Takeuchi}
\email{kat@kaztake.org}
\affiliation{Department of Physics,\! The University of Tokyo,\! 7-3-1 Hongo,\! Bunkyo-ku,\! Tokyo 113-0033,\! Japan}%
\affiliation{Institute for Physics of Intelligence,\! The University of Tokyo,\! 7-3-1 Hongo,\! Bunkyo-ku,\! Tokyo 113-0033,\! Japan}%
\affiliation{Universal Biology Institute,\! The University of Tokyo,\! 7-3-1 Hongo,\! Bunkyo-ku,\! Tokyo 113-0033,\! Japan}%

\date{\today}

\begin{abstract}
    Discovering governing equations, whether manually or by data-driven methods, has been central in physics and related areas.
    Since governing equations are typically constrained by a set of symmetries, using symmetry constraints to restrict terms is usually the first step in manually formulating a governing equation, but it often becomes intractable for complex systems with high-order derivatives or multiple fields.
    When a data-driven method is used, on the other hand, imposing physical constraints such as symmetries typically requires manual preprocessing or computationally expensive iterative procedures.
    Here, we propose an algebraic filtering method that enumerates all symmetry-allowed terms for continuum equations within a finite candidate space.
    By treating symmetry generators as linear operators on the candidate space, we reduce the problem of enforcing both discrete and continuous symmetries to solving a set of linear kernel equations.
    The solution yields a provably complete list of permitted terms.
    We demonstrate the method's effectiveness by identifying invariant terms for systems with dihedral symmetry and recovering the governing equations for the Toner--Tu and Kardar--Parisi--Zhang systems, including higher-order terms useful for extending known models.
    The method provides a systematic way to obtain a symmetry-allowed search space for data-driven equation discovery, e.g., the sparse identification of nonlinear dynamics method.
\end{abstract}

%\keywords{}

\maketitle

% If in two-column mode, this environment will change to single-column
% format so that long equations can be displayed. Use
%\begin{widetext}
%\end{widetext}

\pseudoparagraph{Introduction.}
Inferring and discovering governing equations has been central to many areas of physics and beyond.
Traditionally, governing equations have often been derived by hand based on physical arguments, symmetry principles, and controlled approximations.
However, manual derivation is not always straightforward: multiple fields (e.g., vector fields) or high-order derivatives produce a combinatorially large set of candidate terms, making it difficult to maintain physical consistency.
Recently, automated discovery methods have been developed to derive governing equations from data, including sparse regression \cite{brunton2016sindy,rudy2017pdefind}, symbolic regression \cite{schmidt2009distilling,udrescu2020aifeynman}, and neural-network methods \cite{long2017pdenet,sirignano2018dgm,raissi2019pinns}.
In practice, to be robust against noise, these data-driven methods also require physically informed constraints, most of which are based on symmetries.
Some approaches add symmetry constraints to the loss function \cite{sirignano2018dgm,raissi2019pinns}, but this soft restriction requires careful tuning of penalty weights and often complicates optimization.
Other methods strictly enforce physical constraints by iteratively projecting candidate solutions onto constraint manifolds at each optimization step \cite{kaptanoglu2021project}, resulting in computationally heavy iterations.
Alternatively, discovering equations from a manually curated candidate space \cite{rudy2017pdefind} requires explicitly encoding physical symmetries into the library construction, which suffers from the same combinatorial explosion as manual derivation for systems with multiple fields or high-order derivatives.

In this work, we present an algebraic filtering method that enumerates all symmetry-compatible linear combinations within a chosen finite candidate space, yielding a provably complete list of permitted terms under specified symmetries.
The method treats symmetry generators as linear operators on the candidate space and reduces the invariance conditions to finite kernel problems.
This approach is applicable to both discrete and continuous symmetries.
Applying the method to several classical systems demonstrates its effectiveness in enumerating higher-order terms automatically and exhaustively.

\pseudoparagraph{Theory.}
Here we outline the theoretical framework of our method.
See Supplemental Material \cite{supplemental} for a complete description of the theory and its proof.

We consider systems described by classical fields.
For simplicity, we assume the target system is described by real coordinates $\vec{x}=\ab(x_1, x_2, \ldots, x_d)\in \Omega \subset \Rset^d$ and a real smooth $n$-component field $\ve{u}=\ab(u_1, u_2, \ldots, u_n): \Omega \to \Rset^n$ \footnote{We denote non-spatial vector quantities by bold symbols and spatial vectors by arrowed symbols.}.
We denote the set of all such smooth fields as $\mathcal{U}^n$ where $\mathcal{U}$ is the set of all smooth real functions on $\Omega$.
We further assume that the governing equations depend only on $\vec{x}$ and finitely many derivatives of $\ve{u}$, including the field itself, i.e., we assume the locality of systems.
More specifically, there exists a non-negative integer $K$ and an $m$-component smooth function $\ve{F}= \ab(F_1, F_2, \ldots, F_m) \in \mathcal{F}^m$ such that the equations can be written as
\begin{equation}
    \forall \vec{x} \in \Omega, \ \ve{F}\ab(\vec{x}, \ve{u}\ab(\vec{x}), \vec{\partial}\ve{u}\ab(\vec{x}),\ldots , {\vec{\partial}}^K\ve{u}\ab(\vec{x})) = \ve{0}, \label{eq:govern_long}
\end{equation}
where $\vec{\partial}^k\ve{u}\ab(\vec{x})$ denotes the list of $k$-th order partial derivatives of $\ve{u}$ at point $\vec{x}$.
The set $\mathcal{F}^m$ is the direct product of $\mathcal{F}$, which consists of smooth real functions that take $\vec{x}, \ve{u}\ab(\vec{x}), \vec{\partial}\ve{u}\ab(\vec{x}),\ldots , {\vec{\partial}}^K\ve{u}\ab(\vec{x})$ as arguments.
%This assumption is typically satisfied for systems with locality.
%Regarding the smooth function $\ve{F}$, we consider the set of smooth real functions and smooth $m$-component functions
\Cref{eq:govern_long} defines a solution manifold, denoted by $\mathcal{S}_{\ve{F}} \subset \mathcal{U}^n$. %%% 分の順序がうまくいかない

Under a few assumptions that are usually valid for symmetries considered in physics (see Supplemental Material \cite{supplemental}), we search for governing equations [\cref{eq:govern_long}] that are invariant under the given symmetries.
We call the equations invariant under given symmetries if the solution manifold is invariant under all transformations $\mathcal{T} \in \mathcal{G}$, where $\mathcal{G}$ is the corresponding symmetry group.
Under our definitions, this invariance condition for $\ve{F} \in \mathcal{F}^m$ is expressed as
\begin{equation}
    \forall \mathcal{T} \in \mathcal{G},\  \mathcal{S}_{\ve{F}} =\mathcal{T}\ab(\mathcal{S}_{\ve{F}})=\mathcal{S}_{\mathcal{M}_{\mathcal{T}}\ab(\ve{F})}, \label{eq:invariance_definition_set}
\end{equation}
with a linear map $\mathcal{M}_{\mathcal{T}}$ on $\mathcal{F}$ such that $\mathcal{S}_{\mathcal{M}_{\mathcal{T}}\ab(\ve{F})}=\mathcal{T}\ab(\mathcal{S}_{\ve{F}})$.
We further impose the following covariance condition:
\begin{equation}
    \exists X_\mathcal{T}\in\text{GL}\ab(m,\Rset),\ \mathcal{M}_{\mathcal{T}}\ab(\ve{F}) = X_\mathcal{T} \ve{F}. \label{eq:covariance_F_T}
\end{equation}
This condition guarantees the invariance under the transformation $\mathcal{T}$.

In practice the symmetry group $\mathcal{G}$ is specified by generators.
Let $\{\mathcal{T}_\alpha\}_{\alpha\in A}$ denote discrete generators and let $\mathcal{S}\ab(\ve{\theta})$ denote continuous transformations with multi-dimensional parameter $\ve{\theta} = \ab\{ \theta_\beta\}_{\beta \in B}$ which satisfy $\mathcal{S}\ab(\ve{0})=\mathrm{Id}$ (identity map) \footnote{
    Multiple continuous symmetries can be treated by combining their parameters into a single multi-dimensional parameter $\ve{\theta}$.
}.
Since $\mathcal{M}_{\mathcal{T}}$ forms a representation of the symmetry group $\mathcal{G}$ on $\mathcal{F}$, we can reduce the invariance condition for all $\mathcal{T}\in\mathcal{G}$ to a finite set of linear equations for the generators:
\begin{align}
     & \forall \alpha \in A,\ \exists X_\alpha \in \text{GL}\ab(m,\Rset),\  \mathcal{M}_\alpha \ve{F} = X_\alpha \ve{F}, \label{eq:covariance_reduced_discrete} \\
     & \forall \beta \in B,\ \exists Y_\beta \in \text{End}\ab(m,\Rset),\   \mathcal{D}_\beta \ve{F} = Y_\beta \ve{F},
    \label{eq:covariance_reduced_continuous}
\end{align}
with the covariance condition as a working criterion.
Here, $\mathcal{M}_\alpha$ stands for $\mathcal{M}_{\mathcal{T}_\alpha}$, and
%we denote $\mathcal{M}_{\mathcal{T}_\alpha}$ as $\mathcal{M}_\alpha$ for notational simplicity.
the continuous generators $\ab\{\mathcal{D}_\beta\}_{\beta \in B}$ of the representation on $\mathcal{F}$ are defined by
\begin{equation}
    \mathcal{D}_\beta=\left.\pdv{}{\theta_\beta}\mathcal{M}_{\mathcal{S}\ab(\ve{\theta})}\right|_{\ve{\theta}=\ve{0}}. \label{eq:continuous_generator_definition}
\end{equation}
Detailed proofs of this reduction are provided in Supplemental Material \cite{supplemental}.

While we can reformulate the constraints as generalized eigenproblems, we specialize them for practical use.
Write $\ve{F}=\ve{L}+\ve{R}$ where $\ve{L}$ denotes known (or assumed) terms and $\ve{R}$ denotes unknown terms to be discovered.
We assume that $\ve{L}$ and $\ve{R}$ transform covariantly and are separated under the symmetry action, so that \cref{eq:covariance_reduced_discrete} becomes
\begin{equation}
    \forall \alpha \in A,\ \exists X_\alpha \in \text{GL}\ab(m,\Rset),\
    \begin{cases}
        \mathcal{M}_\alpha \ve{L}=X_\alpha \ve{L}, \\
        \mathcal{M}_\alpha \ve{R}=X_\alpha \ve{R}.
    \end{cases}
    \label{eq:covariance_separated}
\end{equation}

To reduce the problem to a finite set of linear equations, we expand $\ve{L}$ and $\ve{R}$ as
\begin{align}
    L_i & = L^{\tilde{\mu}}_i \tilde{f}_{\tilde{\mu}},\quad L^{\tilde{\mu}}_i \in \Rset, \label{eq:linear_combination_L} \\
    R_i & = R^\mu_i f_\mu,\quad R^\mu_i \in \Rset, \label{eq:linear_combination_R}
\end{align}
where $\ve{f}=\ab(f_1, f_2, \ldots, f_a) \in \mathcal{F}^a$ are the basis functions for the candidate space and $\tilde{\ve{f}}=\ab(\tilde{f}_1, \tilde{f}_2, \ldots, \tilde{f}_c) \in \mathcal{F}^c$ are those for the known terms, which may differ from $f_\mu$.
We refer to $\ve{f}$ and $\tilde{\ve{f}}$ as the candidate basis and analysis basis, respectively.
The action of $\mathcal{M}_\alpha$ on these bases is expanded using additional basis functions $\ve{g}=\ab(g_1, g_2, \ldots, g_b) \in \mathcal{F}^b$ and $\tilde{\ve{g}}=\ab(\tilde{g}_1, \tilde{g}_2, \ldots, \tilde{g}_d) \in \mathcal{F}^d$ as
\begin{align}
    \mathcal{M}_\alpha f_\mu         & = \ab(T_\alpha)_\mu^\nu f_\nu + \ab(D_\alpha)_\mu^\rho \ab(g_\alpha)_\rho, \label{eq:transformed_candidate_term}                                \\
    \mathcal{M}_\alpha \tilde{f}_\mu & = \ab(\tilde{T}_\alpha)_\mu^\nu \tilde{f}_\nu + \ab(\tilde{D}_\alpha)_\mu^\rho \ab(\tilde{g}_\alpha)_\rho. \label{eq:transformed_analysis_term}
\end{align}

Using \cref{eq:covariance_separated,eq:linear_combination_L,eq:linear_combination_R,eq:transformed_candidate_term,eq:transformed_analysis_term} and assuming that the coefficient matrix $L$ for the known terms has full row rank, which is natural since the equations are typically linearly independent, we can finally obtain the following equation as a reduced form of \cref{eq:covariance_reduced_discrete}:
\begin{equation}
    \begin{cases}
        R T_\alpha = L\tilde{T}_\alpha L^+ R, \\
        R D_\alpha = O.
    \end{cases} \label{eq:covariance_constraints_R}
\end{equation}
Here, $L^+$ denotes the Moore-Penrose (right) inverse \cite{wang2018inverse} satisfying $LL^+ = I$ (see Supplemental Material \cite{supplemental}).
This constitutes kernel equations for the unknown coefficients $R$.
For the continuous transformations [\cref{eq:covariance_reduced_continuous}], we can follow the same procedure and obtain the following kernel equations:
\begin{equation}
    \begin{cases}
        R T'_\beta = L\tilde{T}'_\beta L^+ R, \\
        R D'_\beta = O,
    \end{cases} \label{eq:covariance_constraints_R_continuous}
\end{equation}
with
\begin{align}
    \mathcal{D}_\beta f_\mu         & = \ab(T'_\beta)_\mu^\nu f_\nu + \ab(D'_\beta)_\mu^\rho \ab(g'_\beta)_\rho, \label{eq:transformed_term_continuous}                                          \\
    \mathcal{D}_\beta \tilde{f}_\mu & = \ab(\tilde{T}'_\beta)_\mu^\nu \tilde{f}_\nu + \ab(\tilde{D}'_\beta)_\mu^\rho \ab(\tilde{g}'_\beta)_\rho. \label{eq:transformed_analysis_term_continuous}
\end{align}
The only difference is that $L\tilde{T}'_\beta L^+$ need not be invertible.

Thus we have reduced symmetry enforcement to finite kernel equations for the unknown coefficient matrix $R$.
Solving these kernel equations yields a finite-dimensional solution space with arbitrary basis vectors $R_1,R_2,\ldots, R_r$, which correspond to the symmetry-allowed linear combinations of the candidate terms, i.e., the permitted terms $R_1\ve{f},R_2\ve{f},\ldots, R_r\ve{f}$.

\pseudoparagraph{Algorithm.}
The details of the algorithm are presented in \cref{alg:apply_transform,alg:coeff_matrix,alg:transform_matrix,alg:coeff_basis}.
Before explaining the procedures, we first review the key concepts used in the algorithm as inputs, along with their notations and assumptions.

\let\oldReturn\Return
\renewcommand{\Return}{\State\oldReturn}
\begin{figure}
    \begin{algorithm}[H]
        \caption{Apply Transformation to Expressions}\label{alg:apply_transform}
        \begin{algorithmic}[1]
            % \Require $\vec{x}$ is $d$-dimensional coordinate symbols.
            % \Require $\ve{u}\ab(\vec{x})$ is $n$-dimensional field symbols.
            % \Require $\mathcal{T}$ is a transformation with replacement rules for coordinates, fields, and derivatives.
            % \Require $s$ is an expression.
            \Function{apply\_trans}{
            $\ab\{x_i\}_{i=1}^{d}$,
            $\ab\{u_i\}_{i=1}^{n}$,
            $\mathcal{T}$,
            $s$
            }

            \If{$s$ is of the form $f\ab(e_1,e_2,\ldots,e_k)$ where $f:\Rset^k \to \Rset$}
            \For{$i\gets 1,\ldots,k$}
            \State $e'_i \gets$ \Call{apply\_trans}{$\ab\{x_i\}_{i=1}^{d}$, $\ab\{u_i\}_{i=1}^{n}$, $\mathcal{T}$, $e_i$}
            \EndFor
            \Return $f\ab(e'_1,e'_2,\ldots,e'_k)$ \Comment{Recursively apply the transformation to the arguments.}
            \EndIf

            \For{$i\gets 1,\ldots,d$}
            \If{$s$ is a coordinate $x_i$}
            \Return $\ab(\chi_{\mathcal{T}})_i\ab(\vec{x})$ \Comment{Apply coordinate replacement rule.}
            \EndIf
            \EndFor

            \For{$i\gets 1,\ldots,n$}
            \If{$s$ is a field $u_i$}
            \Return $\ab(\ve{v}_{\mathcal{T}})_i\ab(\vec{x}, \ve{u}\ab(\vec{x}), \vec{\partial}\ve{u}\ab(\vec{x}),{\vec{\partial}}^2\ve{u}\ab(\vec{x}),\ldots )$ \Comment{Apply field replacement rule.}
            \EndIf
            \EndFor

            \For{$i\gets 1,\ldots,d$}
            \If{$s$ is a derivative $\pdv{}{x_i}e$}
            \State $e' \gets$ \Call{apply\_trans}{$\ab\{x_i\}_{i=1}^{d}$, $\ab\{u_i\}_{i=1}^{n}$, $\mathcal{T}$, $e$}
            \Return $\ab(\sum_{j=1}^{d} \ab(J_{\mathcal{T}}^{-1})_{i,j}\ab(\vec{x}) \pdv{}{x_j})\ab(e')$ \Comment{Apply derivative replacement rule and recursively apply the transformation to the differentiated term.}
            \EndIf
            \EndFor

            \Return $s$ \Comment{Return original expression if no rule matches.}
            \EndFunction

        \end{algorithmic}
    \end{algorithm}
\end{figure}

\begin{figure}[htbp]
    \begin{algorithm}[H]
        \caption{Compute Coefficient Matrix for Expressions}\label{alg:coeff_matrix}
        \begin{algorithmic}[1]

            \Function{coeff\_matrix}{
            $\ab\{s_i\}_{i=1}^{k}$,
            $\ab\{f_i\}_{i=1}^{l}$
            }
            \For{$i\gets 1,\ldots,k$}
            \State $z_i \gets$ \Call{classify}{$s_i$} \Comment{Expand and simplify $s_i$ into canonical form, then return the dictionary of coefficients for each basis function.}
            \EndFor
            \State $K \gets$ \Call{merge\_sets}{$\ab\{z_i.\text{keys}\}_{i=1}^{k}$}
            \State $K \gets$ \Call{remove\_keys}{$K$, $\ab\{f_i\}_{i=1}^{l}$}
            \State $\ab\{g_i\}_{i=1}^{m} \gets$ \Call{to\_list}{$K$}
            \For{$i\gets 1,\ldots,k$}
            \For{$j\gets 1,\ldots,l$}
            \State $C_{ij} \gets z_i[f_j]$
            \EndFor
            \EndFor

            \For{$i\gets 1,\ldots,k$}
            \For{$j\gets 1,\ldots,m$}
            \State $D_{ij} \gets z_i[g_j]$
            \EndFor
            \EndFor

            \Return $C$, $D$, $\ab\{g_i\}_{i=1}^{m}$

            \EndFunction

        \end{algorithmic}
    \end{algorithm}
\end{figure}

\begin{figure}[htbp]
    \begin{algorithm}[H]
        \caption{Compute Transformation Matrix}\label{alg:transform_matrix}
        \begin{algorithmic}[1]

            \Function{trans\_matrix}{
            $\ab\{x_i\}_{i=1}^{d}$,
            $\ab\{u_i\}_{i=1}^{n}$,
            $\mathcal{T}$,
            $\ab\{f_i\}_{i=1}^{k}$
            }
            \For{$i\gets 1,\ldots,k$}
            \State $f'_i \gets$ \Call{apply\_trans}{$\ab\{x_i\}_{i=1}^{d}$, $\ab\{u_i\}_{i=1}^{n}$, $\mathcal{T}$, $f_i$}
            \EndFor

            \State $T,D,\_ \gets$ \Call{coeff\_matrix}{$\ab\{f'_i\}_{i=1}^{k}$, $\ab\{f_i\}_{i=1}^{k}$}

            \Return $T$, $D$

            \EndFunction

        \end{algorithmic}
    \end{algorithm}
\end{figure}

\begin{figure}[htbp]
    \begin{algorithm}[H]
        \caption{Compute Basis of Solution Space}
        \label{alg:coeff_basis}
        \begin{algorithmic}[1]
            \Procedure{coeff\_basis}{
            $\ab\{x_i\}_{i=1}^{d}$,
            $\ab\{u_i\}_{i=1}^{n}$,
            $\ab\{\mathcal{T}_\alpha\}_{\alpha\in A}$,
            $\ab\{\mathcal{S}_\beta\ab(\theta)\}_{\beta\in B}$,
            $\ab\{f_i\}_{i=1}^{a}$,
            $\ab\{L_i\}_{i=1}^{m}$,
            $\ab\{\tilde{f}_i\}_{i=1}^{c}$
            }

            \State $L, D_L, \_ \gets$ \Call{coeff\_matrix}{$\ab\{L_i\}_{i=1}^{m}$,$\ab\{\tilde{f}_i\}_{i=1}^{c}$}
            \If {$D_L\neq O$ or $\text{rank}\ab(L) < m$}
            \Return \textbf{failure} %\Comment{Known terms are not in the analysis space, fail.}
            \EndIf

            \ForAll{$\alpha \in A$}
            \State $\tilde{T}_\alpha, \tilde{D}_\alpha \gets$ \Call{trans\_matrix}{
            $\ab\{x_i\}_{i=1}^{d}$,
            $\ab\{u_i\}_{i=1}^{n}$,
            $\mathcal{T}_\alpha$,
            $\ab\{\tilde{f}_i\}_{i=1}^{c}$
            }
            \State $X_\alpha \gets L\tilde{T}_\alpha L^+$
            \If{$\text{rank}\ab(X_\alpha) \neq m$ or $L\tilde{D}_\alpha \neq O$}
            \Return \textbf{failure}
            \EndIf

            \State $T_\alpha, D_\alpha \gets$ \Call{trans\_matrix}{
            $\ab\{x_i\}_{i=1}^{d}$,
            $\ab\{u_i\}_{i=1}^{n}$,
            $\mathcal{T}_\alpha$,
            $\ab\{f_i\}_{i=1}^{a}$
            }
            \EndFor

            \ForAll{$\beta \in B$}

            \State $\tilde{T}_\beta, \tilde{D}_\beta \gets$ \Call{trans\_matrix}{
            $\ab\{x_i\}_{i=1}^{d}$,
            $\ab\{u_i\}_{i=1}^{n}$,
            $\mathcal{S}_\beta\ab(\theta)$,
            $\ab\{\tilde{f}_i\}_{i=1}^{c}$
            }
            \State $\tilde{T}'_\beta \gets \pdv{}{\theta}\tilde{T}_\beta$, $\tilde{D}'_\beta \gets \pdv{}{\theta}\tilde{D}_\beta$

            \State $Y_\beta \gets L\tilde{T}'_\beta L^+$
            \If{$L\tilde{D}'_\beta \neq O$}
            \Return \textbf{failure}
            \EndIf

            \State $T_\beta, D_\beta \gets$ \Call{trans\_matrix}{
            $\ab\{x_i\}_{i=1}^{d}$,
            $\ab\{u_i\}_{i=1}^{n}$,
            $\mathcal{S}_\beta\ab(\theta)$,
            $\ab\{f_i\}_{i=1}^{a}$
            }
            \State $T'_\beta \gets \pdv{}{\theta}T_\beta$, $D'_\beta \gets \pdv{}{\theta}D_\beta$
            \EndFor

            \State Solve the kernel equations [\cref{eq:covariance_constraints_R,eq:covariance_constraints_R_continuous}] for all $\alpha\in A$ and $\beta\in B$, then obtain the basis of the solution space $\ab\{R_i\}_{i=1}^{r}$.
            \Return $\ab\{R_i\}_{i=1}^{r}$
            \EndProcedure
        \end{algorithmic}
    \end{algorithm}
\end{figure}

The symbols $\vec{x}=\ab(x_1,\ldots,x_d)=\ab\{x_i\}_{i=1}^{d}$ and $\ve{u}=\ab(u_1,\dots,u_n)=\ab\{u_i\}_{i=1}^{n}$ represent the system's coordinates and fields, respectively.
No specific symmetry constraints are assumed on these symbols \textit{a priori}.
Users may choose any symbolic representation; for example, one can use $x,y,z$ for the coordinates and $u,v,w$ for the fields, or use indexed symbols like $x_i$ and $u_i$.

The symmetry transformations, $\{\mathcal{T}_\alpha\}_{\alpha \in A}$ for discrete symmetries and $\{\mathcal{S}_\beta(\theta)\}_{\beta \in B}$ for continuous symmetries, generate the entire symmetry group.
The discrete transformations $\mathcal{T}_\alpha$ act as standard group generators.
For the continuous transformations $\mathcal{S}_\beta(\theta)$, we consider infinitesimal $\theta$, since it is sufficient to construct the derivatives $\mathcal{D}_\beta$.
This $\mathcal{S}_\beta(\theta)$ amounts to $\mathcal{S}\ab(\theta \ve{e}_\beta)$ with the unit vector $\ve{e}_\beta$ along the $\beta$-th direction.

Each transformation $\mathcal{T}$ is explicitly defined by replacement rules for the new coordinates $\vec{\chi}_{\mathcal{T}}$ and the new fields $\ve{v}_{\mathcal{T}}$.
The replacement rules for derivatives additionally require the inverse Jacobian matrix $J_\mathcal{T}^{-1}$, which can be automatically computed from $\vec{\chi}_{\mathcal{T}}$ as $\ab(J_{\mathcal{T}})_{i,j}=\pdv{\ab(\chi_{\mathcal{T}})_j}{x_i}$.
For continuous transformations $\mathcal{S}_\beta\ab(\theta)$, $\vec{\chi}_{\mathcal{S}_\beta}$, $\ve{v}_{\mathcal{S}_\beta}$ and $J_{\mathcal{S}_\beta}^{-1}$ must be parameterized by $\theta$ such that $\mathcal{S}_\beta\ab(0)=\mathrm{Id}$.
This enables the algorithm to compute the corresponding infinitesimal generator by differentiating these rules at $\theta=0$.
To summarize, the replacement rules are applied as follows:
\begin{itemize}
    \item For coordinates: $\vec{x} \mapsto \vec{\chi}_{\mathcal{T}}\ab(\vec{x})$,
    \item For fields: $\ve{u}\ab(\vec{x}) \mapsto \ve{v}_{\mathcal{T}}\ab(\vec{x}, \ve{u}\ab(\vec{x}), \vec{\partial}\ve{u}\ab(\vec{x}),\ldots )$,
    \item For derivatives: $\pdv{}{x_i} \mapsto \ab(J_{\mathcal{T}}^{-1})_{i,j}\ab(\vec{x}) \pdv{}{x_j}$.
\end{itemize}

The candidate terms $\ve{f}=\ab(f_1,\ldots,f_a)=\ab\{f_i\}_{i=1}^{a}$ constitute the basis for the candidate space.
Users can choose any basis defining the search space for the unknown terms.
A typical choice is to select monomials of coordinates, fields, and their derivatives, restricted by constraints such as maximum derivative order or maximum monomial degree.
Such a basis can be easily generated by a helper function.
For example, \cref{alg:list_monomials} lists all monomial terms up to a maximum order $P_{\text{max}}$.

\begin{figure}
    \begin{algorithm}[H]
        \caption{List All Monomial Terms}\label{alg:list_monomials}
        \begin{algorithmic}[1]

            \Function{list\_monomials}{
            $\ab\{s_i\}_{i=1}^{k}$,
            $\ab\{p_i\}_{i=1}^{k}$,
            $P_{\text{max}}$,
            $P=0$
            }
            \If{$k=0$}
            \Return $\{1\}$
            \EndIf
            \State $L \gets \{\}$
            \For{$i\gets 0,\ldots,\text{floor}\ab(\ab(P_{\text{max}}-P)/p_1)$}
            \State $M \gets$ \Call{list\_monomials}{$\ab\{s_i\}_{i=2}^{k}$, $\ab\{p_i\}_{i=2}^{k}$, $P_{\text{max}}$, $P + i * p_i$}
            \State $\ab\{ N_i \} \gets \ab\{ s_1^i * M_i \}$
            \State $L \gets L + N$
            \EndFor
            \Return $L$
            \EndFunction
        \end{algorithmic}
    \end{algorithm}
\end{figure}

The known terms $\ve{L}=\ab(L_1,\ldots,L_m)=\ab\{L_i\}_{i=1}^{m}$ must be linearly independent and expressible as linear combinations of the analysis basis functions $\tilde{\ve{f}}=\ab(\tilde{f}_1,\ldots,\tilde{f}_c)=\ab\{\tilde{f}_i\}_{i=1}^{c}$, also specified by users.
The transformed known terms must also lie within the span of the analysis basis for $X_\alpha$ and $Y_\beta$ to be computed correctly.
The analysis basis $\tilde{\ve{f}}$ may be identical to $\ve{f}$, or chosen distinctly to optimize performance.

The subroutines of the algorithm proceed as follows.
\cref{alg:apply_transform} applies a transformation $\mathcal{T}$ to an expression $s$ (implementation of $\mathcal{M}_\mathcal{T}$).
\cref{alg:coeff_matrix} extracts the coefficient matrices $C$ and $D$ and additional basis $\{g_i\}_{i=1}^{m}$ for a set of expressions $\{s_i\}_{i=1}^{k}$ with respect to a given basis $\{f_i\}_{i=1}^{l}$.
\cref{alg:transform_matrix} uses \cref{alg:apply_transform,alg:coeff_matrix} to obtain transformation matrices $T, D$ with respect to a given basis $\{f_i\}_{i=1}^{k}$ for a given transformation $\mathcal{T}$, corresponding to \cref{eq:transformed_candidate_term,eq:transformed_analysis_term,eq:transformed_term_continuous,eq:transformed_analysis_term_continuous}.

The main procedure, \Call{coeff\_basis}{} (\cref{alg:coeff_basis}), takes coordinate symbols $\vec{x}$, field symbols $\ve{u}$, symmetry generators defined as replacement rules $\ab\{\mathcal{T}_\alpha\}_{\alpha \in A}$ and $\ab\{\mathcal{S}_\beta\ab(\theta)\}_{\beta\in B}$, candidate terms $\ve{f}$, known terms $\ve{L}$, and analysis basis functions $\tilde{\ve{f}}$.
It first computes the coefficient matrix $L$ of the known term $\ve{L}$ and all transformation matrices: $T_\alpha, D_\alpha, T'_\beta, D'_\beta$ for candidate basis and $\tilde{T}_\alpha, \tilde{D}_\alpha, \tilde{T}'_\beta, \tilde{D}'_\beta$ for the analysis basis.
Then it formulates and solves the kernel equations for the unknown coefficients $R$.
The output $\ab\{R_i\}_{i=1}^{r}$ constitutes a basis for the solution space of the kernel equations.
This basis corresponds to the set of symmetry-allowed linear combinations of the candidate terms, expressed as $R_1\ve{f}, R_2\ve{f}, \ldots, R_r\ve{f}$.

\begin{figure}
    \centering
    \begin{tikzpicture}
        \draw[->,>=stealth] (-2,4) to[bend right=30] node[midway, right] {Rotation} (-3,2);

        \draw[->,>=stealth] (-1,0) to[bend right=30] node[midway, below] {Reflection} (1,0);
        \begin{scope}[x={(-0.8cm,-0.2cm)}, y={(0.8cm,-0.2cm)}, z={(0cm,1cm)}]
            \begin{scope}[xshift=0,yshift=4cm]
                \draw[->,>=stealth] (-0.5,0,0) -- (2, 0, 0);  % x軸
                \draw[->,>=stealth] (0,-0.5,0) -- (0, 2, 0);  % y軸
                \draw[->,>=stealth] (0,0,-0.5) -- (0, 0, 2);  % z軸

                \fill[color=blue, opacity=0.3]
                (1,0,1) -- (2,0,1) -- (1.5,1,0) -- (1,1.5,0) --(0,2,1) -- (-0.5,1,2) -- cycle;

                \draw[->,>=stealth, line width=3, color=red] (0,0,0) -- (1,0,1) node[anchor=south] {$\vec{u}$};
                \draw[->,>=stealth, thick] (0,0,0) -- (2,0,1) node[anchor=south] {$\sin\vec{u}$};
                \draw[->,>=stealth, thick] (0,0,0) -- (1.5,1,0) node[anchor=north] {$\vec{u}^3$};
                \draw[->,>=stealth, thick] (0,0,0) -- (1,1.5,0) node[anchor=north] {$\ab|\vec{u}|^2\vec{u}$};
                \draw[->,>=stealth, thick] (0,0,0) -- (0,2,1) node[anchor=south] {$\vec{\nabla} \times \vec{u}$};
                \draw[->,>=stealth, thick] (0,0,0) -- (-0.5,1,2) node[anchor=south] {$\ab(\vec{\nabla} \cdot \vec{u})\vec{u}$};
            \end{scope}

            \begin{scope}[xshift=-2.5cm,yshift=0]
                \draw[->,>=stealth] (-0.5,0,0) -- (2, 0, 0);  % x軸
                \draw[->,>=stealth] (0,-0.5,0) -- (0, 2, 0);  % y軸
                \draw[->,>=stealth] (0,0,-0.5) -- (0, 0, 2);  % z軸

                \fill[color=blue, opacity=0.3]
                (1,0,1) -- (1,1.5,0) --(0,2,1) -- (-0.5,1,2) -- cycle;

                \draw[->,>=stealth, line width=3, color=red] (0,0,0) -- (1,0,1) node[anchor=south] {$\vec{u}$};
                \draw[->,>=stealth, thick] (0,0,0) -- (1,1.5,0) node[anchor=north] {$\ab|\vec{u}|^2\vec{u}$};
                \draw[->,>=stealth, thick] (0,0,0) -- (-0.5,1,2) node[anchor=south] {$\ab(\vec{\nabla} \cdot \vec{u})\vec{u}$};
                \draw[->,>=stealth, thick] (0,0,0) -- (0,2,1) node[anchor=south] {$\vec{\nabla} \times \vec{u}$};
            \end{scope}

            \begin{scope}[xshift=2.5cm,yshift=0]
                \draw[->,>=stealth] (-0.5,0,0) -- (2, 0, 0);  % x軸
                \draw[->,>=stealth] (0,-0.5,0) -- (0, 2, 0);  % y軸
                \draw[->,>=stealth] (0,0,-0.5) -- (0, 0, 2);  % z軸

                \fill[color=blue, opacity=0.3]
                (1,0,1) -- (1,1.5,0) -- (-0.5,1,2) -- cycle;

                \draw[->,>=stealth, line width=3, color=red] (0,0,0) -- (1,0,1) node[anchor=south] {$\vec{u}$};
                \draw[->,>=stealth, thick] (0,0,0) -- (1,1.5,0) node[anchor=north] {$\ab|\vec{u}|^2\vec{u}$};
                \draw[->,>=stealth, thick] (0,0,0) -- (-0.5,1,2) node[anchor=south] {$\ab(\vec{\nabla} \cdot \vec{u})\vec{u}$};
            \end{scope}

        \end{scope}
    \end{tikzpicture}
    \caption{
        The conceptual visualization of the algorithm: the candidate space is progressively filtered into lower-dimensional subspaces by imposing symmetry constraints such as rotation and reflection.
        \label{fig:algorithm_image}}
\end{figure}
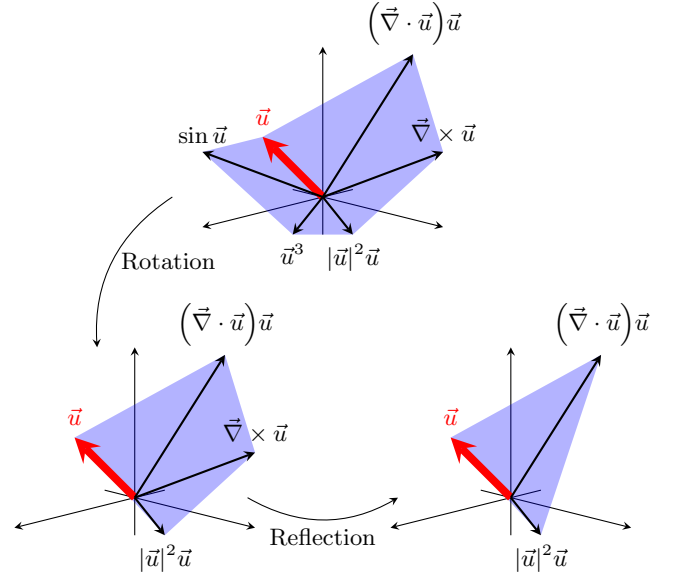

We have implemented this algorithm as a Python library \cite{mylib}, utilizing \texttt{SymPy} \cite{sympy} for symbolic computation.
To enhance performance, the kernel equations are solved sequentially for each transformation.
This iterative process steadily reduces the dimension of the solution space basis, significantly lowering the computational cost of subsequent symbolic operations.
This suggests a filter-like interpretation, as depicted in \cref{fig:algorithm_image}, where symmetry constraints act as filters to sieve symmetry-allowed terms from the candidate space.

\pseudoparagraph{Demonstrations.}
We demonstrate the validity and utility of our method through three examples.
First, we use the dihedral group $D_n$ as a benchmark to verify that the algorithm correctly recovers the known invariant polynomials.
We then apply the method to the systems considered by the Toner--Tu equation \cite{toner1995flocks,toner1998flocking,toner2012reanalysis} and the Kardar--Parisi--Zhang (KPZ) equation \cite{kardar1986kpz}.
As described below, we successfully enumerate all symmetry-allowed terms systematically, which include not only terms occasionally overlooked in manual derivations, but also higher-order contributions.

\pseudosubparagraph{Example 1: $D_n$-invariant polynomials.}
We first consider a simple example of discrete rotation and reflection symmetries in two-dimensional space, i.e., the dihedral group $D_n$.
Let the coordinates be $\vec{v}=\ab(x,y)$ and the field be a scalar $u:\Rset^2 \to \Rset$.
Converting to polar coordinates $\ab(r,\theta)$, the discrete rotation by angle $\frac{2\pi}{N}$ and reflection are represented as the shift and the inversion of $\theta$:
\begin{align}
    \mathcal{R} & :\ \ab(r,\theta) \mapsto \ab(r,\theta + \frac{2\pi}{N}),\ u \mapsto u, \label{eq:dn_rotation} \\
    \mathcal{X} & :\ \ab(r,\theta) \mapsto \ab(r,-\theta),\ u \mapsto u. \label{eq:dn_reflection}
\end{align}
Using Fourier series expansion in $\theta$, the field $u$ with invariance under these transformations is expressed as
\begin{equation}
    u\ab(r,\theta) = \sum_{k=0}^\infty A_k\ab(r) \cos\ab(Nk \theta). \label{eq:dn_candidate}
\end{equation}
Further, restricting $u$ to be a polynomial in $x$ and $y$, we obtain the expanded representation $u = \sum_{k=0}^\infty \sum_{l=0}^\infty B_{k,l} f_{k,l}$, where the basis polynomials are defined as
\begin{equation}
    f_{k,l} := \ab(x^2+y^2)^k \mathrm{Re}\ab\ab((x+iy)^{lN}\ab) = r^{2k+lN}\cos\ab(lN\theta).
    \label{eq:dn_polynomial_basis}
\end{equation}

To apply our algorithm, we specifically consider the $D_5$ group ($N=5$) and search for invariant polynomials up to a total degree of $10$ by setting the constant $1$ as the known term $\ve{L}$.
The resulting basis $R_1\ve{f}, R_2\ve{f}, \ldots, R_r\ve{f}$ for the invariant polynomials is shown in \cref{tab:dn_result}.
\begin{table}[b]
    \caption{\label{tab:dn_result}List of invariant polynomial terms under the discrete rotation by angle $\frac{2\pi}{5}$ and reflection, up to the total degree $10$. The left column shows the raw outputs of the algorithm, while the right column expresses them as linear combinations of the natural basis $f_{k,l}$ defined in \cref{eq:dn_polynomial_basis}.}
    \begin{ruledtabular}
        \begin{tabular}{cc}
            \textbf{raw output $R_i\ve{f}$}                                                                        & \textbf{natural basis}                    \\
            \hline
            $1$                                                                                                    & $f_{0,0}$                                 \\
            $x^{2} + y^{2}$                                                                                        & $f_{1,0}$                                 \\
            $x^{4} + 2 x^{2} y^{2} + y^{4}$                                                                        & $f_{2,0}$                                 \\
            $x^{6} + 3 x^{4} y^{2} + 3 x^{2} y^{4} + y^{6}$                                                        & $f_{3,0}$                                 \\
            $x^{8} + 4 x^{6} y^{2} + 6 x^{4} y^{4} + 4 x^{2} y^{6} + y^{8}$                                        & $f_{4,0}$                                 \\
            $\frac{4 x^{10}}{5} + 9 x^{8} y^{2} - 12 x^{6} y^{4} + 30 x^{4} y^{6} + y^{10}$                        & $\frac{9}{10}f_{5,0}-\frac{1}{10}f_{0,2}$ \\
            $\frac{x^{5}}{5} - 2 x^{3} y^{2} + x y^{4}$                                                            & $\frac15 f_{0,1}$                         \\
            $\frac{x^{7}}{5} - \frac{9 x^{5} y^{2}}{5} - x^{3} y^{4} + x y^{6}$                                    & $\frac15 f_{1,1}$                         \\
            $\frac{x^{9}}{5} - \frac{8 x^{7} y^{2}}{5} - \frac{14 x^{5} y^{4}}{5} + x y^{8}$                       & $\frac15 f_{2,1}$                         \\
            $\frac{x^{10}}{25} - \frac{4 x^{8} y^{2}}{5} + \frac{22 x^{6} y^{4}}{5} - 4 x^{4} y^{6} + x^{2} y^{8}$ & $\frac{1}{50}f_{5,0}+\frac{1}{50}f_{0,2}$
        \end{tabular}
    \end{ruledtabular}
\end{table}
This result perfectly matches \cref{eq:dn_polynomial_basis} up to the total degree $10$, confirming that our method can exhaustively enumerate symmetry-invariant terms.
Note that the basis obtained by the algorithm is not always a natural basis.
For example, the 6th and 10th terms in \cref{tab:dn_result} are linear combinations of $f_{5,0}$ and $f_{0,2}$.

In the following examples, we only show the terms re-expressed in natural forms.

\pseudosubparagraph{Example 2: Toner--Tu equation.}
As a non-trivial physical example, we consider the Toner--Tu equation \cite{toner1995flocks,toner1998flocking,toner2012reanalysis} describing the collective motion of self-propelled particles:
\begin{equation}
    \begin{split}\label{eq:tonertu}
        \pdv{\vec{v}}{t} & + \lambda_1 \ab(\vec{v}\cdot\vec{\nabla})\vec{v} + \lambda_2 \ab(\vec{\nabla}\cdot\vec{v})\vec{v} + \lambda_3 \vec{\nabla}\ab\|\vec{v}\|^2 \\
                         & = \alpha \vec{v} - \beta \ab\|\vec{v}\|^2 \vec{v}                                                                                          \\
                         & + D_1 \vec{\nabla}\ab(\vec{\nabla}\cdot\vec{v}) + D_T \vec{\nabla}^2 \vec{v} + D_2 \ab(\vec{v} \cdot \vec{\nabla})^2 \vec{v}.
    \end{split}
\end{equation}
Here we omitted pressure terms.
While the equation was originally proposed in \cite{toner1995flocks}, the terms $\lambda_2 \ab(\vec{\nabla}\cdot\vec{v})\vec{v}$ and $\lambda_3 \vec{\nabla}\ab\|\vec{v}\|^2$ were initially not present and added in the later analyses \cite{toner1998flocking,toner2012reanalysis}.
Our method provides a systematic way to construct such equations.
By treating the symmetry generators algebraically, we can ensure that all permitted terms are included.

To demonstrate this, we searched for covariant terms using the velocity field $\vec{v}:\Rset^3 \to \Rset^3$ and coordinates $\vec{x}=\ab(x,y,z)$, under rotation and reflection symmetries.
The candidate space consisted of polynomials of $\vec{v}$ and its spatial derivatives up to third order in fields and second order in derivatives (e.g., $v_y v_z \pdv{v_x}{x}$).
Our algorithm successfully recovered all terms of the complete Toner--Tu equation, as shown in \cref{tab:tonertu_partial_result}.
\begin{table}[b]
    \caption{\label{tab:tonertu_partial_result} List of Toner--Tu covariant terms with $\ve{v}$ under rotation and reflection symmetries, up to third order in fields and second order in derivatives. The first group contains terms appearing in \cref{eq:tonertu}. while the second group lists symmetry-allowed terms not present in \cref{eq:tonertu}.}

    \begin{ruledtabular}
        \begin{tabular}{c}
            \textbf{terms}                                                                  \\
            \hline
            $\vec{v}$                                                                       \\
            $\ab\|\vec{v}\|^2 \vec{v}$                                                      \\
            $\vec{\nabla} \ab\|\vec{v}\|^2$                                                 \\
            $\ab(\vec{v}\cdot\vec{\nabla})\vec{v}$                                          \\
            $\ab(\vec{\nabla}\cdot \vec{v})\vec{v}$                                         \\
            $\vec{\nabla}^2 \vec{v}$                                                        \\
            $\vec{\nabla}\ab(\vec{\nabla} \cdot \vec{v})$                                   \\
            $\ab(\vec{v}\cdot\vec{\nabla})^2 \vec{v}$                                       \\

            % u
            % ((u∙u)u)
            % (∇(u∙u))
            % ((u∙∇)u)
            % ((∇∙u)u)
            % ((∇^2)u)
            % (∇(∇∙u))
            % ((u∙∇)((u∙∇)u))

            \hline

            % not listed but hand made

            $\vec{\nabla}\ab(\vec{\nabla}\cdot\ab(\ab\|\vec{v}\|^2\vec{v}))$                \\% (∇(∇∙((u∙u)u)))
            $\vec{\nabla}\ab(\vec{v}\cdot\ab(\vec{\nabla}\ab\|\vec{v}\|^2))$                \\% (∇(u∙(∇(u∙u))))
            $\vec{\nabla}\times\ab(\vec{\nabla}\times\ab(\ab\|\vec{v}\|^2\vec{v}))$         \\% (∇×(∇×((u∙u)u)))
            $\vec{\nabla}\times\ab(\vec{v}\times\ab(\vec{\nabla}\ab\|\vec{v}\|^2))$         \\% (∇×(u×(∇(u∙u))))
            $\vec{\nabla}\times\ab(\vec{v}\times\ab(\ab(\vec{v}\cdot\vec{\nabla})\vec{v}))$ \\% (∇×(u×((u∙∇)u)))
            $\ab(\vec{\nabla}\cdot\vec{v})\ab(\vec{\nabla}\ab\|\vec{v}\|^2)$                \\% ((∇∙u)(∇(u∙u)))
            $\ab(\vec{\nabla}\cdot\vec{v})\ab(\ab(\vec{v}\cdot\vec{\nabla})\vec{v})$        \\% ((∇∙u)((u∙∇)u))
            $\ab(\vec{\nabla}\cdot\vec{v})\ab(\ab(\vec{\nabla}\cdot\vec{v})\vec{v})$        \\% ((∇∙u)((∇∙u)u))
            $\ab(\vec{\nabla}\cdot\ab(\vec{\nabla}\ab\|\vec{v}\|^2))\vec{v}$                \\% ((∇∙(∇(u∙u)))u)
            $\ab(\vec{\nabla}\cdot\ab(\ab(\vec{v}\cdot\vec{\nabla})\vec{v}))\vec{v}$        \\% ((∇∙((u∙∇)u))u)
            $\ab(\vec{\nabla}\cdot\ab(\ab(\vec{\nabla}\cdot\vec{v})\vec{v}))\vec{v}$        \\% ((∇∙((∇∙u)u))u)
            $\ab\|\vec{v}\|^2\ab(\nabla^2\vec{v})$                                          \\% ((u∙u)((∇^2)u))
            $\ab(\vec{v}\cdot\ab(\nabla^2\vec{v}))\vec{v}$                                  \\% ((u∙((∇^2)u))u)
            $\vec{v}\times\ab(\vec{\nabla}\times\ab(\ab(\vec{v}\cdot\vec{\nabla})\vec{v}))$   % (u×(∇×((u∙∇)u)))
        \end{tabular}
    \end{ruledtabular}
\end{table}
Beyond the standard terms, we identified 14 additional symmetry-allowed terms at the same order (second order in derivatives, third order in fields).
We verified that all these terms can be constructed from $\vec{v}$ and $\vec{\nabla}$ using standard vector operations, such as $(\nabla(\vec{v}\cdot(\nabla\times\vec{v})))\times \vec{v}$.
While many of these may be irrelevant in the renormalization-group sense, their systematic enumeration ensures that no symmetry-compatible contribution is missed before performing scaling analysis.

\pseudosubparagraph{Example 3: KPZ equation.}
The power of this algebraic approach is even more evident when dealing with symmetries that involve field-dependent transformations.
A prime example is the statistical tilt symmetry of the KPZ equation, which describes surface growth and other phenomena \cite{kardar1986kpz} (for a recent review, see, e.g., \cite{takeuchi201877kpz}).
The KPZ equation reads:
\begin{equation}
    \pdv{h}{t} = \nu \vec{\nabla}^2 h + \frac{\lambda}{2} \ab(\vec{\nabla} h)^2 + \eta, \label{eq:kpz_equation}
\end{equation}
where $\eta$ is white Gaussian noise.
This system possesses spatial rotation, reflection, and translation symmetries.
For field-dependent symmetries, it has a global shift symmetry $h \mapsto h + h_0$ and the statistical tilt symmetry, the latter of which represents an invariance under the following infinitesimal transformation:
\begin{equation}
    \vec{x} \mapsto \vec{x}- \lambda \vec{\epsilon} t,
    t       \mapsto t,
    h       \mapsto h+\vec{\epsilon}\cdot\vec{x}.
    \label{eq:kpz_statistical_tilt}
\end{equation}
Our method handles this symmetry by using its generator, even without an explicit finite transformation.

Specifically, we searched for invariant terms up to fifth order in fields and fourth order in derivatives.
The results, summarized in \cref{tab:kpz_result}, include all terms of the original KPZ equation [\cref{eq:kpz_equation}] and several higher-order terms.
Notably, the algorithm identified complex terms involving time derivatives and higher-order spatial derivatives that are required to maintain the statistical tilt symmetry.
The same set of terms was obtained for both $d=2$ and $d=3$ spatial dimensions.
\begin{table}
    \caption{\label{tab:kpz_result} KPZ invariant terms with $\lambda=1$ under rotation, reflection, shift, and statistical-tilt symmetries, up to fifth order in fields and fourth order in derivatives.}
    \begin{ruledtabular}
        \begin{tabular}{c}
            \textbf{terms}                                                \\
            \hline
            $1$                                                           \\
            $\vec{\nabla}^2 h$                                            \\
            $\vec{\nabla}^4 h$                                            \\
            $-2\pdv{h}{t}+\ab(\vec{\nabla} h)^2$                          \\
            $\ab(\vec{\nabla}^2 h)^2$                                     \\
            $\ab(- 2 \pdv{h}{t} + \ab(\vec{\nabla} h)^2)\vec{\nabla}^2 h$ \\
            $\ab(-2\pdv{h}{t}+\ab(\vec{\nabla} h)^2)^2 $                  \\
            $\vec{\nabla}^2\ab(- 2 \pdv{h}{t} + \ab(\vec{\nabla} h)^2)$   \\
            $\ab\|\vec{\nabla}\vec{\nabla} h\|^2$                         \\
            $\pdv[2]{h}{t}
                -2 \ab(\vec{\nabla} h)\cdot \ab(\vec{\nabla} \pdv{h}{t})
                +\ab(\vec{\nabla} h)\cdot\ab(\vec{\nabla}\vec{\nabla} h)\cdot\ab(\vec{\nabla} h)$
        \end{tabular}
    \end{ruledtabular}
\end{table}

\pseudoparagraph{Conclusions and Outlook.}
In this work, we have developed a systematic algebraic filtering method to identify all continuum equation terms permitted by a given symmetry group.
By treating symmetry transformations as linear operators on a finite-dimensional candidate space, we reduced the problem of symmetry enforcement to solving a set of kernel equations.
This approach provides a provably complete list of invariant or covariant terms within the chosen basis, effectively eliminating the risk of omitting physically relevant terms due to human error.
Our demonstrations on the Toner--Tu and KPZ equations confirm the practical utility of the method.
Besides confirming all the known terms, our method successfully listed higher-order terms permitted by the symmetries considered in each case, providing a solid basis for constructing extended models.

There are several directions for future work.
First, the symmetry-filtered libraries generated by our method can be directly integrated into data-driven discovery frameworks, especially sparse regression, such as the sparse identification of nonlinear dynamics (SINDy) \cite{brunton2016sindy}.
By restricting the search space to symmetry-consistent terms, these methods can achieve greater robustness against noise and ensure the physical validity of the discovered equations.
Second, while our current implementation relies on symbolic manipulation, improving algorithmic performance through numerical linear-algebra solvers or more efficient basis representations will be essential for handling substantially larger candidate libraries.
Finally, extending this framework to accommodate more complex systems, such as those considered in quantum field theory, would broaden its applicability to a wider range of fundamental physical systems.

\begin{acknowledgments}
    \pseudoparagraph{Acknowledgments.}
    This work is supported in part by JSPS KAKENHI (Grant Numbers JP23K17664 and JP24K00593), Japan Science and Technology Agency (JST) FOREST (Grant Number JPMJFR2364), and the JSPS Core-to-Core Program ``Advanced core-to-core network for the physics of self-organizing active matter'' (JPJSCCA20230002).
    J.Y. acknowledges support from FoPM, WINGS Program, the University of Tokyo.
\end{acknowledgments}

\bibliography{ref}

\end{document}

\title{Supplementary Material for ``Symmetry-Informed Term Filtering for Continuum Equation Discovery''}

\author{Junya Yokokura}
\email{junya-yk@g.ecc.u-tokyo.ac.jp}
\affiliation{Department of Physics,\! The University of Tokyo,\! 7-3-1 Hongo,\! Bunkyo-ku,\! Tokyo 113-0033,\! Japan}%

\author{Kazumasa A. Takeuchi}
\email{kat@kaztake.org}
\affiliation{Department of Physics,\! The University of Tokyo,\! 7-3-1 Hongo,\! Bunkyo-ku,\! Tokyo 113-0033,\! Japan}%
\affiliation{Institute for Physics of Intelligence,\! The University of Tokyo,\! 7-3-1 Hongo,\! Bunkyo-ku,\! Tokyo 113-0033,\! Japan}%
\affiliation{Universal Biology Institute,\! The University of Tokyo,\! 7-3-1 Hongo,\! Bunkyo-ku,\! Tokyo 113-0033,\! Japan}%

\date{\today}

\maketitle

\section*{Full Description of Theory}

We consider systems described by classical fields.
For simplicity, we assume the target system is described by real coordinates $\vec{x}=\ab(x_1, x_2, \ldots, x_d)\in \Omega \subset \Rset^d$ and a real smooth $n$-component field $\ve{u}=\ab(u_1, u_2, \ldots, u_n): \Omega \to \Rset^n$ \footnote{We denote non-spatial vector quantities by bold symbols and spatial vectors by arrowed symbols.}.
We denote the set of all such smooth fields as $\mathcal{U}^n$ where $\mathcal{U}$ is the set of all smooth real functions on $\Omega$.
We further assume that the governing equations depend only on $\vec{x}$ and finitely many derivatives of $\ve{u}$, including the field itself.
More precisely, there exists a non-negative integer $K$ and an $m$-component smooth function $\ve{F}= \ab(F_1, F_2, \ldots, F_m)$ such that the equations can be written as
\begin{equation}
    \forall \vec{x} \in \Omega, \ \ve{F}\ab(\vec{x}, \ve{u}\ab(\vec{x}), \vec{\partial}\ve{u}\ab(\vec{x}),\ldots , {\vec{\partial}}^K\ve{u}\ab(\vec{x})) = \ve{0}, \label{eq:govern_long}
\end{equation}
where $\vec{\partial}^k\ve{u}\ab(\vec{x})$ denotes the list of $k$-th order partial derivatives of $\ve{u}$ at point $\vec{x}$.
This assumption is typically satisfied for systems with locality.

For notational simplicity we use jet bundles \cite{saunders1989jet}: the $k$-th order jet of $\ve{u}$ at $\vec{x}$ is denoted by $j^k_{\vec{x}}\ve{u}$ and the bundle of all such jets by $\mathcal{J}^k\ab(\Omega,\Rset^n)$.
Practically, the jet $j^k_{\vec{x}}\ve{u}$ can be treated as the tuple formed by the coordinates $\vec{x}$, the field values $\ve{u}\ab(\vec{x})$, and all partial derivatives of $\ve{u}$ at $\vec{x}$ up to order $k$.
The bundle $\mathcal{J}^k\ab(\Omega,\Rset^n)$ can be treated as the set of all such tuples for all $\vec{x}\in\Omega$ and all smooth fields $\ve{u}:\Omega \to \Rset^n$.
Let $\mathcal{F}$ denote the set of smooth functions on $\mathcal{J}^k\ab(\Omega,\Rset^n)$ and $\mathcal{F}^m$ denote the set of $m$-component functions on $\mathcal{J}^k\ab(\Omega,\Rset^n)$.
Especially, each component $F_i$ is an element of $\mathcal{F}$ and the entire function $\ve{F}$ is an element of $\mathcal{F}^m$.
We can then write \cref{eq:govern_long} in the compact form
\begin{equation}
    \forall \vec{x} \in \Omega, \ \ve{F}\ab(j^K_{\vec{x}} \ve{u}) = \ve{0}. \label{eq:govern}
\end{equation}
These equations define a solution manifold, denoted by $\mathcal{S}_{\ve{F}} \subset \mathcal{U}^n$.

Then we define the class of symmetries under consideration.
A symmetry transformation $\mathcal{T}: \mathcal{U}\to\mathcal{U}$ is assumed to satisfy the following property: for a transformed field $\ve{u}_{\mathcal{T}}'=\mathcal{T}\ab(\ve{u})$, its value $\ve{u}'_{\mathcal{T}}\ab(\vec{x}')$ at transformed coordinates $\vec{x}'$ can be expressed in terms of the original coordinates $\vec{x}$ and finitely many derivatives of $\ve{u}$ at $\vec{x}$, including the field itself.
More precisely, we assume the existence of a smooth bijection of coordinates $\vec{\chi}_{\mathcal{T}}:\Omega \to \Omega$, a non-negative integer $L$, and a smooth function $\ve{v}_{\mathcal{T}}: \mathcal{J}^L\ab(\Omega, \Rset^n) \to \Rset^n$ such that
\begin{equation}
    \forall \vec{x} \in \Omega, \ \ve{u}'_{\mathcal{T}}\ab(\vec{\chi}_{\mathcal{T}}\ab(\vec{x})) = \ve{v}_{\mathcal{T}}\ab(j^L_{\vec{x}} \ve{u}), \label{eq:trans}
\end{equation}
or equivalently,
\begin{equation}
    \forall \vec{x} \in \Omega, \ \ve{u}'_{\mathcal{T}}\ab(\vec{x}) = \ve{v}_{\mathcal{T}}\ab(j^L_{\vec{\chi}_{\mathcal{T}}^{-1}\ab(\vec{x})} \ve{u}). \label{eq:trans_inverse}
\end{equation}
This assumption holds for symmetries that preserve the locality of the system.

With these assumptions, we search for governing equations [\cref{eq:govern}] that are invariant under the given symmetries.
We call the equations invariant under given symmetries if the solution manifold is invariant under all transformations $\mathcal{T} \in \mathcal{G}$, where $\mathcal{G}$ is the corresponding symmetry group.
Under our definitions, this invariance condition for $\ve{F} \in \mathcal{F}^m$ is expressed as
\begin{equation}
    \forall \mathcal{T} \in \mathcal{G},\  \mathcal{S}_{\ve{F}} =\mathcal{T}\ab(\mathcal{S}_{\ve{F}}), \label{eq:invariance_definition_set}
\end{equation}
or equivalently,
\begin{equation}
    \begin{split}
         & \forall \mathcal{T} \in \mathcal{G},\ \forall \ve{u} \in \mathcal{U},\
        \ab(\forall \vec{x} \in \Omega, \ \ve{F}\ab(j^K_{\vec{x}} \ve{u}) = \ve{0}) \\
         & \Leftrightarrow
        \ab(\forall \vec{x} \in \Omega, \ \ve{F}\ab(j^K_{\vec{x}} \ve{u}'_{\mathcal{T}}) = \ve{0}).
        \label{eq:invariance_definition_expanded}
    \end{split}
\end{equation}

Here, considering \cref{eq:trans_inverse}, the jet of the transformed field $j^K_{\vec{x}} \ve{u}'_{\mathcal{T}}$ can be constructed from that of the original field by applying the chain rule to each derivative in $j^K_{\vec{x}}$ and propagating it through $\ve{v}_{\mathcal{T}}$.
Because this procedure increases the necessary derivative order by at most $L$, we can express this mapping with a function $y_{\mathcal{T}}: \mathcal{J}^{K+L}\ab(\Omega, \Rset^n) \to \mathcal{J}^{K}\ab(\Omega, \Rset^n)$ as
\begin{equation}
    \forall \vec{x} \in \Omega, \ j^K_{\vec{x}} \ve{u}'_{\mathcal{T}} = y_{\mathcal{T}}\ab(j^{K+L}_{\vec{\chi}_{\mathcal{T}}^{-1}\ab(\vec{x})} \ve{u}). \label{eq:govern_transformed}
\end{equation}
The transformed equations may be written as
\begin{equation}
    \forall \vec{x} \in \Omega, \ \ve{F}\ab(y_{\mathcal{T}}\ab(j^{K+L}_{\vec{\chi}_{\mathcal{T}}^{-1}\ab(\vec{x})} \ve{u})) = \ve{0}, \label{eq:govern_transformed_expanded_nonlocal}
\end{equation}
or equivalently,
\begin{equation}
    \forall \vec{x} \in \Omega, \ \ve{F}'_{\mathcal{T}} \ab(j^{K+L}_{\vec{x}} \ve{u}) = \ve{0}, \label{eq:govern_transformed_expanded_local}
\end{equation}
where $\ve{F}'_{\mathcal{T}} =\ve{F} \circ y_{\mathcal{T}}$.
Here we used the bijection property of $\vec{\chi}_{\mathcal{T}}$.
\Cref{eq:govern_transformed_expanded_local} is more convenient since $\ve{F}'_{\mathcal{T}}$ depends on the same arguments as $\ve{F}$ except for the order of derivatives.
To simplify the later discussions, we extend $\mathcal{F}$ to the set of all smooth functions defined on $J^K\ab(\Omega, \Rset^n)$ for any $K\in\Zset_{\ge0}$, so that $\ve{F}'_{\mathcal{T}}$ remains an element of $\mathcal{F}^m$.
This modification does not affect the preceding discussions, as long as the domains of the other functions are extended accordingly.

Thus the invariance condition [\cref{eq:invariance_definition_expanded}] reduces to the requirement that, for every $\mathcal{T}\in\mathcal{G}$ and every field $\ve{u}\in\mathcal{U}^n$, \cref{eq:govern} and \cref{eq:govern_transformed_expanded_local} are equivalent.
While this equivalence is not trivial in general, it holds in typical cases when the transformed governing functions are related to the original ones by an invertible linear map.
Concretely, suppose there exists an $m\times m$ invertible matrix $X_\mathcal{T}\in\text{GL}\ab(m,\Rset)$ such that
\begin{equation}
    \ve{F}'_{\mathcal{T}} =X_\mathcal{T} \ve{F}, \label{eq:covariance_F_T}
\end{equation}
then the invariance condition is proven as
\begin{equation}
    \begin{split}
        \ab(\Rightarrow)\  & \forall \vec{x} \in \Omega, \ \ve{F}'_{\mathcal{T}}\ab(j^{K+L}_{\vec{x}} \ve{u}) = X_\mathcal{T} \ve{F}\ab(j^K_{\vec{x}} \ve{u}) = X_\mathcal{T}\ve{0} = \ve{0}, \\
        \ab(\Leftarrow)\   & \forall \vec{x} \in \Omega, \ \ve{F}\ab(j^K_{\vec{x}} \ve{u}) = X_\mathcal{T}^{-1} \ve{F}'_{\mathcal{T}}\ab(j^{K+L}_{\vec{x}} \ve{u}) = \ve{0}.
    \end{split}\label{eq:invariance_proved_by_covariance}
\end{equation}
We refer to \cref{eq:covariance_F_T} as the covariance condition on $\ve{F}$ under $\mathcal{T}$ and adopt it as the working criterion for invariance.
This choice is natural in practice because $X_\mathcal{T}$ plays the role of the representation matrix of the symmetry action on the space of governing functions.

We can reformulate this process from the perspective of representations of symmetry groups.
Each transformation $\mathcal{T}$ derives a map $\mathcal{M}_\mathcal{T}: \mathcal{F} \to \mathcal{F}$, which acts on $F \in \mathcal{F}$ as
\begin{equation}
    \mathcal{M}_\mathcal{T}\ab(F) := F \circ y_{\mathcal{T}}, \label{eq:representation_definition}
\end{equation}
as implicitly used in \cref{eq:govern_transformed_expanded_local}.
Since $\mathcal{F}$ is a vector space over $\Rset$ and $\mathcal{M}_\mathcal{T}$ is a linear map, we can show that $\ab\{\mathcal{M}_\mathcal{T}\}_{\mathcal{T}\in\mathcal{G}}$ is a representation of the symmetry group $\mathcal{G}$ on $\mathcal{F}$ by verifying the homomorphism property:
\begin{equation}
    \begin{split}
        \mathcal{M}_{\mathcal{T}_1\mathcal{T}_2}\ab(F) & =F\circ y_{\mathcal{T}_1\mathcal{T}_2} = F \circ y_{\mathcal{T}_1}\circ y_{\mathcal{T}_2} \\
                                                       & =\mathcal{M}_{\mathcal{T}_2}\ab(\mathcal{M}_{\mathcal{T}_1}\ab(F)),                       \\
        \mathcal{M}_{\mathcal{T}_1\mathcal{T}_2}       & =\mathcal{M}_{\mathcal{T}_2} \mathcal{M}_{\mathcal{T}_1}.
    \end{split} \label{eq:F_representation_homomorphism}
\end{equation}
Note that the order of the composition is reversed.

As demonstrated in \cref{eq:invariance_proved_by_covariance}, two sets of equations are equivalent if their components span the same linear subspace in $\mathcal{F}$.
Thus, instead of working with an explicit list of equations, it is more convenient to reformulate \cref{eq:covariance_F_T} as the invariance of the spanned subspace $\ab<\ve{F}>:=\text{span}\ab\{F_1,F_2,\ldots,F_m\}$:
\begin{equation}
    \mathcal{M}_\mathcal{T}\ab<\ve{F}> = \ab<\ve{F}>. \label{eq:covariance_F_T_space}
\end{equation}
This invariance property derives a subgroup $\mathcal{G}_{\ve{F}} \subset \mathcal{G}$ as
\begin{equation}
    \mathcal{G}_{\ve{F}} := \ab\{\mathcal{T}\in \mathcal{G}: \mathcal{M}_\mathcal{T}\ab<\ve{F}> = \ab<\ve{F}> \}. \label{eq:subgroup_covariance}
\end{equation}
The conjunction of all covariance conditions on $\ve{F}$ is expressed with $\mathcal{G}_{\ve{F}}$ as
\begin{equation}
    \mathcal{G}_{\ve{F}} = \mathcal{G}. \label{eq:covariance_group}
\end{equation}

In practice the symmetry group $\mathcal{G}$ is specified by generators.
Let $\{\mathcal{T}_\alpha\}_{\alpha\in A}$ denote discrete generators and let $\mathcal{S}\ab(\ve{\theta})$ denote continuous transformations with multi-dimensional parameter $\ve{\theta} = \ab\{ \theta_\beta\}_{\beta \in B}$ which satisfy $\mathcal{S}\ab(\ve{0})=\mathrm{Id}$ (identity map) \footnote{
    Multiple continuous symmetries can be treated by combining their parameters into a single multi-dimensional parameter $\ve{\theta}$.
}.
Since the subgroup containing these transformations constitutes the entire group $\mathcal{G}$, \cref{eq:covariance_group} is equivalent to
\begin{align}
     & \forall \alpha \in A,\ \mathcal{T}_\alpha \in \mathcal{G}_{\ve{F}}, \label{eq:covariance_discrete_generators}            \\
     & \exists U \in \mathcal{N}\ab(\ve{0}),\ \forall \ve{\theta} \in U,\ \mathcal{S}\ab(\ve{\theta}) \in \mathcal{G}_{\ve{F}},
    \label{eq:covariance_continuous_neighborhood}
\end{align}
where $\mathcal{N}\ab(\ve{\theta})$ is a set of neighborhoods of $\ve{\theta}$.

For continuous transformations, we can further simplify the condition.
Let us denote the generators of the representation on $\mathcal{F}$ by $\ab\{\mathcal{D}_\beta\}_{\beta \in B}$ as
\begin{equation}
    \mathcal{D}_\beta=\left.\pdv{}{\theta_\beta}\mathcal{M}_{\mathcal{S}\ab(\ve{\theta})}\right|_{\ve{\theta}=\ve{0}}. \label{eq:continuous_generator_definition}
\end{equation}
Then we can select a neighborhood $U \in \mathcal{N}\ab(\ve{0})$ such that
\begin{equation}
    \forall \ve{\theta} \in U,\ \exists \mathcal{D} \in \text{span}\ab\{\mathcal{D}_\beta\}_{\beta\in B},\ \mathcal{M}_{\mathcal{S}\ab(\ve{\theta})} = e^\mathcal{D}. \label{eq:exponential_mapping}
\end{equation}
With \cref{eq:exponential_mapping} we can show that \cref{eq:covariance_continuous_neighborhood} is equivalent to
\begin{equation}
    \forall \beta \in B,\ \mathcal{D}_\beta\ab<\ve{F}> \subset \ab<\ve{F}>. \label{eq:covariance_continuous_generators}
\end{equation}
Obtaining \cref{eq:covariance_continuous_generators} from \cref{eq:covariance_continuous_neighborhood} is straightforward by expanding the derivative in \cref{eq:continuous_generator_definition}.
Conversely, to obtain \cref{eq:covariance_continuous_neighborhood} from \cref{eq:covariance_continuous_generators}, we use the Taylor expansion of the exponential mapping $e^\mathcal{D}$.
This expansion, together with the finite-dimensionality of $\ab<\ve{F}>$, shows $\mathcal{M}_{\mathcal{S}\ab(\ve{\theta})}\ab<\ve{F}> = e^\mathcal{D}\ab<\ve{F}> \subset \ab<\ve{F}>$ for all $\ve{\theta} \in U$.
Moreover, since $\mathcal{S}\ab(\ve{0})=\mathrm{Id}$, the continuity of the rank with respect to $\ve{\theta}$ ensures that the mapping preserves the dimension of $\ab<\ve{F}>$ within a sufficiently small neighborhood $U' \subset U$.
Consequently, we obtain $\mathcal{M}_{\mathcal{S}\ab(\ve{\theta})}\ab<\ve{F}> = \ab<\ve{F}>$ for any $\ve{\theta} \in U'$, completing the proof.

As a result, checking the invariance condition [\cref{eq:invariance_definition_set}] reduces to
\begin{align}
     & \forall \alpha \in A,\ \exists X_\alpha \in \text{GL}\ab(m,\Rset),\  \mathcal{M}_\alpha \ve{F} = X_\alpha \ve{F}, \label{eq:covariance_reduced_discrete} \\
     & \forall \beta \in B,\ \exists Y_\beta \in \text{End}\ab(m,\Rset),\   \mathcal{D}_\beta \ve{F} = Y_\beta \ve{F},
    \label{eq:covariance_reduced_continuous}
\end{align}
with covariance conditions as working criterion.
Here and in the following, we denote $\mathcal{M}_{\mathcal{T}_\alpha}$ as $\mathcal{M}_\alpha$ for notational simplicity.

Although the covariance conditions reduce invariance checking to linear-algebra constraints, solving them for arbitrary functions in $\mathcal{F}^m$ remains infeasible.
We therefore restrict attention to a finite-dimensional candidate subspace $\ab<\ve{f}>$, where $\ve{f}=\ab(f_1,f_2,\ldots,f_a)\in\mathcal{F}^a$ is a chosen basis of the space.
We call $\ab<\ve{f}>$ the candidate space and refer to $f_\mu$ as candidate terms.
In this setting each component of $\ve{F}$ is represented as a linear combination:
\begin{equation}
    F_i = C^\mu_i f_\mu,\quad C^\mu_i \in \Rset. \label{eq:linear_combination_simple}
\end{equation}
Here and in the following, we use Einstein summation notation for repeated indices.

We first address the discrete-generator constraints in \cref{eq:covariance_reduced_discrete}.
The transformed image $\mathcal{M}_\alpha f_\mu$ may not lie in the candidate space, so we introduce overflow basis functions $\ve{g}_\alpha=\ab((g_\alpha)_1,(g_\alpha)_2,\ldots,(g_\alpha)_{b_\alpha})\in\mathcal{F}^{b_\alpha}$ to complete the representation.
This allows us to represent the transformed candidate terms as
\begin{equation}
    \mathcal{M}_\alpha f_\mu = \ab(T_\alpha)_\mu^\nu f_\nu + \ab(D_\alpha)_\mu^\rho \ab(g_\alpha)_\rho, \label{eq:transformed_candidate_term}
\end{equation}
where $\ab(T_\alpha)_\mu^\nu, \ab(D_\alpha)_\mu^\rho \in \Rset$ are the transformation matrices for the candidate terms and the overflow basis functions, respectively.
Hence
\begin{equation}
    \mathcal{M}_\alpha F_i = C^\mu_i\mathcal{M}_\alpha f_\mu = C^\mu_i\ab((T_\alpha)_\mu^\nu f_\nu + (D_\alpha)_\mu^\rho (g_\alpha)_\rho).\label{eq:transformed_equation}
\end{equation}
Comparing this with the covariance requirement yields linear-algebra constraints on the coefficients $C^\mu_i$:
\begin{equation}
    \begin{split}
         & \exists X_\alpha \in \text{GL}\ab(m,\Rset), \\ &C^\mu_i \ab(\ab(T_\alpha)_\mu^\nu f_\nu + \ab(D_\alpha)_\mu^\rho \ab(g_\alpha)_\rho) = \ab(X_\alpha)_i^j C^\mu_j f_\mu. \label{eq:covariance_algebraic}
    \end{split}
\end{equation}
Comparing the coefficients of $f_\mu$ and $\ab(g_\alpha)_\rho$ on both sides, we obtain
\begin{equation}
    \exists X_\alpha \in \text{GL}\ab(m,\Rset),\ \begin{cases}
        C^\mu_i \ab(T_\alpha)_\mu^\nu = \ab(X_\alpha)_i^j C^\nu_j, \\
        C^\mu_i \ab(D_\alpha)_\mu^\rho = 0,
    \end{cases} \label{eq:covariance_constraints}
\end{equation}
or using matrix notation,
\begin{equation}
    \exists X_\alpha \in \text{GL}\ab(m,\Rset),\ \begin{cases}
        CT_\alpha = X_\alpha C, \\
        CD_\alpha = O,
    \end{cases} \label{eq:covariance_constraints_matrix}
\end{equation}
These are the linear-algebra constraints on the coefficients $C^\mu_i$.

While we can reformulate the constraints as generalized eigenproblems, we specialize them for practical use.
Write $\ve{F}=\ve{L}+\ve{R}$ where $\ve{L}$ denotes known (or assumed) terms and $\ve{R}$ denotes unknown terms to be discovered.
We assume that $\ve{L}$ and $\ve{R}$ transform covariantly and are separated under the symmetry action, so that \cref{eq:covariance_reduced_discrete} becomes
\begin{equation}
    \forall \alpha \in A,\ \exists X_\alpha \in \text{GL}\ab(m,\Rset),\
    \begin{cases}
        \mathcal{M}_\alpha \ve{L}=X_\alpha \ve{L}, \\
        \mathcal{M}_\alpha \ve{R}=X_\alpha \ve{R}.
    \end{cases}
    \label{eq:covariance_separated}
\end{equation}
We expand $\ve{L}$ and $\ve{R}$ as
\begin{align}
    L_i & = L^{\tilde{\mu}}_i \tilde{f}_{\tilde{\mu}},\quad L^{\tilde{\mu}}_i \in \Rset, \label{eq:linear_combination_L} \\
    R_i & = R^\mu_i f_\mu,\quad R^\mu_i \in \Rset. \label{eq:linear_combination_R}
\end{align}
where $\tilde{\ve{f}}=\ab(\tilde{f}_1, \tilde{f}_2, \ldots, \tilde{f}_c) \in \mathcal{F}^c$ are the basis functions for known terms, which can be different from $f_\mu$.
Let us call them the analysis basis functions.
They provide transformation matrices different from those of candidate terms as
\begin{equation}
    \mathcal{M}_\alpha \tilde{f}_\mu = \ab(\tilde{T}_\alpha)_\mu^\nu \tilde{f}_\nu + \ab(\tilde{D}_\alpha)_\mu^\rho \ab(\tilde{g}_\alpha)_\rho. \label{eq:transformed_analysis_term}
\end{equation}
Following the previous process, \cref{eq:covariance_separated} is rewritten as
\begin{equation}
    \exists X_\alpha \in \text{GL}\ab(m,\Rset),\ \begin{cases}
        L\tilde{T}_\alpha = X_\alpha L, \\
        L\tilde{D}_\alpha = O,          \\
        RT_\alpha = X_\alpha R,         \\
        RD_\alpha = O.
    \end{cases} \label{eq:covariance_constraints_matrix_separated}
\end{equation}

To proceed, we assume the coefficient matrix $L$ for the known terms has full row rank, which is natural since the equations are typically linearly independent.
In that case $X_\alpha$ can be computed as $X_\alpha=L\tilde{T}_\alpha L^+$, where $L^+$ denotes the Moore--Penrose (right) inverse \cite{wang2018inverse} satisfying $LL^+=I$.
The constraints on the unknown coefficients $R$ then read
\begin{equation}
    \begin{cases}
        R T_\alpha = L\tilde{T}_\alpha L^+ R, \\
        R D_\alpha = O,
    \end{cases} \label{eq:covariance_constraints_R}
\end{equation}
which are kernel equations for the unknown coefficients $R$.
Note that $L\tilde{T}_\alpha L^+$ is not necessarily invertible in general, so one should verify $L\tilde{T}_\alpha L^+\in\text{GL}\ab(m,\Rset)$ holds since the covariance condition requires an invertible representation.

For the continuous transformations [\cref{eq:covariance_reduced_continuous}], we can follow the same procedure and obtain the following kernel equations:
\begin{equation}
    \begin{cases}
        R T'_\beta = L\tilde{T}'_\beta L^+ R, \\
        R D'_\beta = O,
    \end{cases} \label{eq:covariance_constraints_R_continuous}
\end{equation}
with
\begin{align}
    \mathcal{D}_\beta f_\mu         & = \ab(T'_\beta)_\mu^\nu f_\nu + \ab(D'_\beta)_\mu^\rho \ab(g'_\beta)_\rho, \label{eq:transformed_term_continuous}                                          \\
    \mathcal{D}_\beta \tilde{f}_\mu & = \ab(\tilde{T}'_\beta)_\mu^\nu \tilde{f}_\nu + \ab(\tilde{D}'_\beta)_\mu^\rho \ab(\tilde{g}'_\beta)_\rho. \label{eq:transformed_analysis_term_continuous}
\end{align}
The only difference is that $L\tilde{T}'_\beta L^+$ need not be invertible.

Thus we have reduced symmetry enforcement to finite kernel equations for the unknown coefficient matrix $R$.
Solving these kernel equations yields a finite-dimensional solution space with arbitrary basis vectors $R_1,R_2,\ldots, R_r$, which correspond to the symmetry-allowed linear combinations of the candidate terms, i.e., the permitted terms $R_1\ve{f},R_2\ve{f},\ldots, R_r\ve{f}$.

%\begin{thebibliography}{99}
%\end{thebibliography}
\bibliography{ref}